\documentclass[reprint,showpacs,amsmath,amssymb,aps,prb,superscriptaddress]{revtex4-1}

\usepackage{graphicx}
\usepackage{dcolumn}
\usepackage{bm}

\begin{document}

\title{A brief review of thermal transport in mesoscopic systems from nonequilibrium Green's function approach}

\author{Zhizhou Yu}
\author{Guohuan Xiong}
\author{Lifa Zhang}\email{phyzlf@njnu.edu.cn}
\affiliation{NNU-SULI Thermal Energy Research Center (NSTER) $\&$ Center for Quantum Transport and Thermal Energy Science (CQTES), School of Physics and Technology, Nanjing Normal University, Nanjing 210023, China}

\begin{abstract}
With the rapidly increasing integration density and power density in nanoscale electronic devices, the thermal management concerning heat generation and energy harvesting becomes quite crucial. Since phonon is the major heat carrier in semiconductors, thermal transport due to phonons in mesoscopic systems has attracted much attention. In quantum transport studies, the nonequilibrium Green's function (NEGF) method is a versatile and powerful tool that has been developed for several decades. In this review, we will discuss theoretical investigations of thermal transport using the NEGF approach from two aspects. For the aspect of phonon transport, the phonon NEGF method is briefly introduced and its applications on thermal transport in mesoscopic systems including one-dimensional atomic chains, multi-terminal systems, and transient phonon transport are discussed. For the aspect of thermoelectric transport, the caloritronic effects in which the charge, spin, and valley degrees of freedom are manipulated by the temperature gradient are discussed. The time-dependent thermoelectric behavior is also presented in the transient regime within the partitioned scheme based on the NEGF method.
\end{abstract}


\maketitle

\section{Introduction}\label{sec1}
As transistor gate lengths are scaled down into the 10-nm regime with the rapid development of nanotechnology, millions of transistors are fabricated within a square millimeter in the integrated circuit chip\cite{Pop1705144}. With the increasing transistor density in chips, the power density raises rapidly, which becomes the roadblock for the continued miniaturization of integrated circuits since the enhanced chip temperature prevents the reliable performance of integrated circuits. In order to design next-generation devices with low energy consumption, it is crucial to study the thermal transport in nanostructures to understand heat generation and dissipation. Recently, numerous researchers have proposed various theoretical models to study the fundamental physics in thermal transport and carried out experiments on low-dimensional nanomaterials to show their potential applications in thermal engineering \cite{RMP_Li,Zhang_2017,ChenXB,Cahill}.

Phonon, the physical quasiparticle representing the mechanical vibrations, is responsible for the transmission of heat in solids. Understanding and controlling the transport properties of phonons provide opportunities to reduce heat consumption and utilize waste heat. Various prototypical phononic devices such as thermal diodes\cite{Li184301}, thermal transistors\cite{Li143501,Chung054402}, thermal logic gates\cite{Li177208}, and thermal memories \cite{Li267203} have been proposed to manipulate the heat flow at the nanoscale. Recently, the chirality of phonons has been observed experimentally in monolayer tungsten diselenide \cite{Zhu579}. The discovery of chiral phonons has received wide attention in emerging fields such as valleytronics \cite{Lu093901,Lu3999} and topological states \cite{Liu064106}. Therefore, exploring the mechanisms of phonon transport and scattering in nanoscale phononic devices is of great importance for artificially tuning thermal transport properties for future heat management in electronic devices and specific applications in phononic devices.

Apart from phonon transport, the thermoelectric effect which describes a direct conversion from heat energy to electric energy and vice versa, is another major concern in the field of thermal transport due to its potential applications in harvesting and recovering heat. The performance of thermoelectric materials at a certain temperature is evaluated by the dimensionless figure of merit ($ZT$). The big challenge lying behind the thermoelectric technology is the improvement of $ZT$ value of thermoelectric materials, namely, simultaneous enhancement in the electrical conductivity and reduction in the lattice thermal conductivity \cite{TWAHA2016698,Li2150}. In the past decades, the thermoelectric behavior of a series of low-dimensional materials has been theoretically predicted and experimentally studied, which exhibits huge potential in the application of high-performance thermoelectric devices \cite{Zhao13184,Lee12011,Chang778,Babaei}. However, it is still an open question and a long way to search for better thermoelectric materials and further improve the $ZT$ value.

The method of nonequilibrium Green's function (NEGF) is a versatile and powerful tool to study both electronic and phononic transport properties in nanoscale materials. The NEGF method was used to investigate quantum electric transport by Caroli et al. for the first time in 1971\cite{Caroli_1971}. An explicit formula for the transmission coefficient and tunneling current was derived in terms of the Green's function. A Landauer formula for the current through an interacting electron region was derived by Meir and Wingreen, which provided a modern framework to study the electronic transport in mesoscopic systems \cite{Meir}. The general formula of time-dependent electric current through the interacting and noninteracting mesoscopic systems was derived using the Keldysh NEGF technique \cite{Jauho5528}. Besides the electronic transport, the NEGF method was used to treat the phonon transport in solid junctions by Wang et al. and the formula of thermal current due to atomic vibrations was presented in terms of Green's function\cite{NEGF3,NEGF4}. Within the NEGF approach, many-body effects in quantum transport such as electron-phonon and electron-electron interactions can be included through self-energies without deviating the framework \cite{Sergueev146803,Shimazaki075110,Paulsson201101,Ferretti116802,Thygesen115333}. The NEGF method was also combined with the density functional theory (DFT) which is an art-of-the-state technique for modeling and predicting the electronic transport properties of nanomaterials \cite{Taylor245407,Brandbyge165401}.

In this review, we aim to give a brief summary of theoretical studies on thermal transport including the phonon and thermoelectric transport in mesoscopic systems by using the NEGF method. In Sec.~\ref{sec2}, we first introduce the phonon NEGF method and its applications on thermal transport. The interfacial thermal transport in one-dimensional atomic chains, phonon transport in multi-terminal systems, and time-dependent phonon transport in the transient regime are discussed. In Sec.~\ref{sec3}, the basic concepts of thermoelectricity are introduced. The dc thermoelectric transport and its application on spin and valley caloritronics are discussed within the linear response theory. The time-dependent thermoelectric transport in the transient regime within the partitioned scheme was also presented. Finally, a brief conclusion and outlook are given in Sec.~\ref{sec4}.

\section{Phonon transport}\label{sec2}
\subsection{NEGF method for phonon transport}
Various methods have been used to study the phonon transport, such as molecular dynamics (MD) \cite{MD1,MD2,MD3,RMP_Li}and Boltzmann transport equation (BTE) method \cite{Boltzman1,Boltzman2,Boltzman3}. The MD method can incorporate nonlinearity. However, it is only valid at high temperatures and becomes not accurate at low temperatures due to its classical nature. The BTE method is usually used to study the thermal conductivities for bulk materials and can not be used for systems without translational invariance. For the mesoscopic system in which quantum effects dominate the phonon transport, NEGF is an effective approach in a whole diffusive to ballistic regime\cite{NEGF1, NEGF2, NEGF3, NEGF4}. In this section, we first give a quick review of the NEGF technique in phononic systems.

\subsubsection{phonon current}
We consider a nonconducting solid that only the vibrational degrees of freedom are treated. The Hamiltonian is given by\cite{NEGF3,NEGF4},
\begin{equation}\label{H}
  H = \sum_{\alpha = L,C,R} H_\alpha + u_L^\dag V_{LC} u_C + u_C^\dag V_{CR} u_R,
\end{equation}
where $L,C,R$ denotes the left lead, central region, and right lead, respectively.
\begin{equation}\label{Halpha}
  H_\alpha = \frac{1}{2}\dot{u}_\alpha^\dag \dot{u}_\alpha + \frac{1}{2} u_\alpha^\dag K_\alpha u_\alpha,
\end{equation}
where $u_\alpha$ is the column vector consisting of all displacement variables in region $\alpha$ and $\dot{u}_\alpha$ is the corresponding conjugate momentum. $K_\alpha$ is the spring constant matrix. $V_{CL} = V^\dag_{LC}$ and $V_{CR} = V^\dag_{RC}$ are the coupling matrices of the central region to the left and right leads, respectively. The dynamic matrix for a full linear system can be written as,
\begin{equation}\label{K}
  K = \left(
        \begin{array}{ccc}
          K_L    & V_{LC} & 0      \\
          V_{CL} & K_{C}  & V_{CR} \\
          0      & V_{RC} & K_{R} \\
        \end{array}
      \right).
\end{equation}

The phonon current flow from the left lead to the central region can be defined as \cite{NEGF3,NEGF4},
\begin{equation}\label{IL}
  J_L = - \langle \dot{H}_L(t)\rangle.
\end{equation}
By using the Heisenberg equation of motion, we can obtain
\begin{equation}\label{I2}
  J_L = \langle \dot{u}_L^\dag(t) V_{LC} u_{C}(t)\rangle.
\end{equation}
By defining the following lesser Green's function \cite{NEGF3,NEGF4},
\begin{equation}\label{GCL}
  G^<_{CL}(t,t') = -i \langle u_L(t') u_C(t)^T \rangle^T,
\end{equation}
we have
\begin{equation}\label{I3}
  J_L = i\frac{\partial}{\partial t'} \mathrm{Tr}[G^<_{CL}(t,t')V^{LC}]\Big|_{t=t'}.
\end{equation}
After the Fourier transformation, the current can be expressed as,
\begin{equation}\label{IG}
  J_L = - \int^{+\infty}_{-\infty} \frac{d\omega}{2\pi} \omega \mathrm{Tr} [ G^<_{CL} (\omega) V_{LC} ].
\end{equation}
In order to solve $G^<_{CL} (\omega)$ in the above equation, we can relate $G_{CC}$ to $G_{CL}$ by using the Dyson equation\cite{Jauhobook},
\begin{equation}\label{Dyson}
  G_{CL}(\tau, \tau_1) = \int_{C} d\tau_2 G_{CC}(\tau, \tau_2) V^{CL} g_L(\tau_2, \tau_1) .
\end{equation}
Here, the integral is along the contour. $g_L$ is the contour-ordered Green's function for the isolated left lead in equilibrium. By employing the analytic continuation, we can obtain\cite{Jauhobook}
\begin{equation}\label{GCL2}
  G^<_{CL}(\omega) = G^r_{CC}(\omega) V_{CL} g^<_L(\omega) +  G^<_{CC}(\omega) V_{CL} g^a_L(\omega).
\end{equation}
Substituting Eq.~(\ref{GCL2}) into Eq.~(\ref{IG}), the expression of the phonon current becomes
\begin{equation}\label{II}
  J_L = - \int^{+\infty}_{-\infty} \frac{d\omega}{2\pi}  \omega \mathrm{Tr} [G^r(\omega) \Pi^<_L(\omega) +  G^<(\omega) \Pi^a_L(\omega)],
\end{equation}
where $\Pi_L^{\gamma} = V_{CL} g^{\gamma}_L V_{LC}$ ($\gamma = r,a,<,>$) is the self-energy due to the interaction with leads. By taking $(J_L+J^*_L)/2$, we can obtain \cite{NEGF3,NEGF4}
\begin{eqnarray}\label{I2}
  J_L &=& - \int^{+\infty}_{-\infty} \frac{d\omega}{4\pi}  \omega \mathrm{Tr} \big\{[G^r(\omega)-G^a(\omega)] \Pi^<_L(\omega) \nonumber\\
  &&+  G^<(\omega) [\Pi^a_L(\omega) - \Pi^r_L(\omega)]\big\},
\end{eqnarray}
which can be further written in the form of Meir-Wingreen formula \cite{Meir,NEGF2}
\begin{equation}\label{IMW}
  J_L = - \int^{+\infty}_{-\infty}\frac{d\omega}{4\pi}  \omega \mathrm{Tr} [G^<(\omega) \Pi^>_L(\omega) -  G^>(\omega) \Pi^<_L(\omega)].
\end{equation}

The retarded Green's function in the frequency domain for a steady-state transport can be written as \cite{NEGF3,NEGF4},
\begin{equation}\label{Gr}
  G^r = [(\omega + i0^+)^2 \mathbb{I} - K^C - \Pi^r]^{-1}.
\end{equation}
Here, $\mathbb{I}$ represents the identity matrix and $0^+$ is an infinitesimal positive number. The lesser Green's function satisfies the Keldysh equation\cite{Jauhobook},
\begin{equation}\label{keldysh}
  G^<  = G^r\Pi^< G^a.
\end{equation}

Now we introduce the phonon bandwidth function,
\begin{equation}\label{Gamma}
  \Lambda_\alpha = i(\Pi^r_{\alpha} - \Pi^a_{\alpha}).
\end{equation}
then Eq.~(\ref{I2}) can be expressed as the Landauer formula\cite{Mingo,NEGF2},
\begin{equation}\label{Ifinal}
  J_L = \int^{+\infty}_{0} \frac{d\omega}{2\pi} \omega \Xi(\omega) (n_{L} - n_{R}),
\end{equation}
where $n_{\alpha}(\omega) = 1/[\exp(\omega/T_\alpha)-1]$ is the Bose-Einstein distribution function in lead $\alpha$ ($k_B$ = 1 for simplicity) and
\begin{equation}\label{Transmission}
  \Xi(\omega) = \mathrm{Tr} (G^r \Lambda_L G^a \Lambda_R),
\end{equation}
is the phonon transmission coefficient in the form of the Caroli formula\cite{Caroli_1971}. More details of the basic definition and properties of phonon NEGF can be found in Refs.~\onlinecite{NEGF1,NEGF2}.

We define the phonon thermal conductance as \cite{NEGF4},
\begin{equation}\label{kph1}
 \kappa_{ph} = \lim_{\Delta T \rightarrow 0} \frac{J}{\Delta T},
\end{equation}
where $\Delta T$ is the temperature difference of two leads. For ballistic transport, the phonon conductance can be expressed in the form of Landauer-like formula \cite{Yamamoto}
\begin{equation}\label{kph}
 \kappa_{ph} = \int^{+\infty}_{0} \frac{d\omega}{2\pi} \omega \Xi(\omega) \frac{\partial n}{\partial T}.
\end{equation}

\subsubsection{Nonlinear systems}
In the following, we discuss the quantum self-consistent mean-field theory based on the NEGF method to deal with nonlinear thermal transport. We introduce the quartic interaction term into the Hamiltonian as an example, which can be given by \cite{NEGF3}
\begin{equation}\label{Hn}
  H_n = \frac{1}{4} \sum_{ijkl} T_{ijkl} u_{C,i}u_{C,k}u_{C,j}u_{C,l}.
\end{equation}
We can also handle the cubic interaction term for the thermal transport. By applying the equation of motion, the Green's function with the nonlinearity can be written as\cite{Zhang_2013},
\begin{eqnarray}\label{GF1}
&& \frac{\partial^2}{\partial \tau^2} G_{im}(\tau,\tau_1) + \sum_j K_{C,ij} G_{jm} (\tau,\tau_1) \nonumber\\
&& + \sum_{jkl} T_{ijkl} G_{jklm} (\tau,\tau,\tau,\tau_1) \nonumber\\
&=& -\delta(\tau - \tau_1) \delta_{im} - \sum_j \int d\tau_2 \Pi_{ij}(\tau,\tau_2) G_{jm}(\tau_2,\tau_1), \nonumber\\
\end{eqnarray}
where $G(\tau_1,\tau_2,\tau_3,\tau_4) = -i \langle T_c u(\tau_1)u(\tau_2)u(\tau_3)u(\tau_4)\rangle$ with $T_c$ the time-order operator is the four-point Green's function. Within the mean-field approximation, the four-point Green's function can be represented by the two-point Green's function \cite{NEGF2},
\begin{eqnarray}\label{GF2}
  -iG(\tau_1,\tau_2,\tau_3,\tau_4) & \approx & G(\tau_1,\tau_2)G(\tau_3,\tau_4) + G(\tau_1,\tau_3)G(\tau_2,\tau_4) \nonumber\\
   && +G(\tau_1,\tau_4)G(\tau_2,\tau_3).
\end{eqnarray}
Then we can obtain
\begin{eqnarray}\label{GF1}
&& \frac{\partial^2}{\partial \tau^2} G_{im}(\tau,\tau_1) + \sum_j K_{C,ij} G_{jm} (\tau,\tau_1) \nonumber\\
&& + 3i\sum_{jkl} T_{ijkl} G_{kl}(0) G_{jm}(\tau, \tau_1) \nonumber\\
&=& -\delta(\tau - \tau_1) \delta_{im} - \sum_j \int d\tau_2 \Pi_{ij}(\tau,\tau_2) G_{jm}(\tau_2,\tau_1). \nonumber\\
\end{eqnarray}
Therefore, we can account the nonlinearity by the following self-energy,
\begin{equation}\label{Signman}
  \Pi_{n,ij} = 3i\sum_{kl} T_{ijkl} G_{kl}(0) = 3\sum_{kl} T_{ijkl} \langle u_k u_l\rangle,
\end{equation}
where
\begin{equation}\label{Signman}
  \langle u_k u_l\rangle = i \int^{\infty}_0 \frac{d\omega}{2\pi} G^<(\omega).
\end{equation}
We note that this nonlinear self-energy is real and it only shifts the frequencies of phonon modes.

By introducing the nonlinear self-energy, the retarded Green's function with nonlinearity can be then written as
\begin{equation}\label{Grn}
  G^r = [(\omega + i0^+)^2 \mathbb{I} - K^C - \Pi^r - \Pi_n]^{-1}.
\end{equation}
With the help of the Keldysh equation, i.e., Eq.~(\ref{keldysh}), the retarded Green's function can be solved self-consistently. Since we are considering an effectively harmonic problem, the phonon current can be still calculated from Eq.~(\ref{Ifinal}).

\subsubsection{Electron-phonon interaction}
For thermal transport through the metal-semiconductor interface, energy must transfer between electrons and phonons. Therefore, it is highly desirable to understand the heat dissipation for thermal transport through the interface with the electron-phonon interaction. The electron-phonon coupling in the central region can be described by \cite{Jauhobook}
\begin{equation}\label{Heph}
  H_C^{eph} = \sum_i \epsilon_i d_i^\dag d_i + \sum_{ijk} M^k_{ij} d_i^\dag d_j u_k,
\end{equation}
where $d_i^\dag$($d_i$) is the electron creation (annihilation) operator and $\epsilon_i$ is the electron energy level in the central region. $M^k_{ij}$ is the electron-phonon coupling matrix element. Since the Landauer formula of phonon current is only applicable to quasi-ballistic transport, one needs to use the Meir-Wingree formula, i.e., Eq.~(\ref{IMW}), to calculate the phonon current of inelastic processes, including the electron-phonon scattering.

The electron-phonon interaction is included as a perturbation. The full retarded Green's function within the electron-phonon coupling can be obtained from the Dyson equation \cite{Jauhobook},
\begin{equation}\label{Dysonep}
  \bar{G}^r = G^r + G^r \Pi^r_{eph} \bar{G}^r,
\end{equation}
where $G^r$ given in Eq.~(\ref{Gr}) is the bare phonon retarded Green's function without electron-phonon interaction. The Keldysh equation for the system with electron-phonon coupling becomes
\begin{equation}\label{keldyshep}
  \bar{G}^<  = \bar{G}^r(\Pi^< + \Pi^<_{eph})\bar{G}^a.
\end{equation}

Under the Born approximation, we can obtain the nonlinear self-energy due to the electron-phonon interaction up to the second order \cite{Lu_ep,Zhang_ep},
\begin{equation}\label{lesserep}
  \Pi^<_{eph,mn} = -i M^m_{lk} \int \frac{d\epsilon}{2\pi} G^<_{ki} (\epsilon) G^>_{jl}(\epsilon-\omega) M^n_{ij},
\end{equation}
and
\begin{eqnarray}\label{retardedep}
  \Pi^r_{eph, mn} &=& -i M^m_{lk} \int \frac{d\epsilon}{2\pi} [G^r_{ki} (\epsilon) G^<_{jl}(\epsilon-\omega) \nonumber\\
  && + G^<_{ki}(\epsilon) G^a_{jl}(\epsilon-\omega)  ]M^n_{ij}.
\end{eqnarray}

From Eqs.~(\ref{Dysonep})-(\ref{retardedep}), we perform iterations under the self-consistent Born approximation by replacing the bare Green's function $G$ with the full Green's function $\bar{G}$. Finally, the phonon energy current with the electron-phonon interaction can be given in the form of Meir-Wingreen formula \cite{NEGF4,Lu_ep},
\begin{equation}\label{IMWep}
  J_L = - \int^{+\infty}_{-\infty} \frac{d\omega}{4\pi}  \omega \mathrm{Tr} [\bar{G}^<(\omega) \Pi^>_L(\omega) -  \bar{G}^>(\omega) \Pi^<_L(\omega)].
\end{equation}

\subsection{Interfacial thermal transport in one-dimensional atomic chains}
In thermal transport, the interfacial thermal scattering becomes extremely important as the dimension of thermal devices shrinks to the nanoscale. In low dimensional system, it was found that the interfaces can dramatically affect the thermal transport\cite{Gordiz2015,Chenjie2015,Cahill}. In recent years, interfacial thermal transport has been extensively studied by both classical and quantum approaches. To study the interfacial thermal transport, the most widely used models are the acoustic mismatch model \cite{Little} and the diffuse mismatch model \cite{Swartz}. However, both models are lack of accuracy in calculating the interfacial thermal resistance since they neglect the atomic details of actual interface structures. The NEGF approach, which is a powerful method to treat nonequilibrium and interacting systems, has been extensively applied to study interfacial thermal transport. Moreover, the NEGF method can offer a straightforward way to treat nonlinear systems.

\begin{figure}
  \includegraphics[width=3.5in]{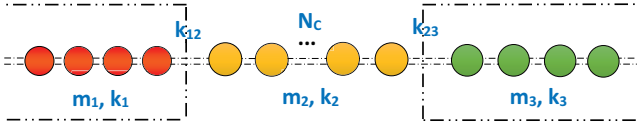}\\
  \caption{Schematic of the one-dimensional atomic chain model. The central region is coupled with the left and right leads by harmonic springs with constant strength $k_{12}$ and $k_{23}$, respectively. The left lead, central region, and right lead are all harmonic chains with the spring constant and mass $k_1$, $m_1$, $k_2$, $m_2$, $k_3$, $m_3$, respectively.}
  \label{schematic}
\end{figure}

One-dimensional atomic chain model has been extensively used to study the interfacial thermal transport, which can provide fundamental physical pictures for practical thermal devices. The one-dimensional atomic chain consists of two semi-infinite leads and a central region, as shown in Fig.~\ref{schematic}. The left and right leads are in equilibrium at different temperatures $T_L$ and $T_R$, respectively. The central region is coupled with the left and right leads by harmonic springs with constant strength $k_{12}$ and $k_{23}$, respectively. The left lead, central region, and right lead are all harmonic chains with the spring constant and mass $k_1$, $m_1$, $k_2$, $m_2$, $k_3$, $m_3$, respectively. The total Hamiltonian of the one-dimensional atomic chain can be given by \cite{Zhang_2011}
\begin{equation}\label{H_1Dchain}
  H = \sum_{\alpha=L,C,R} H_\alpha + \frac{1}{2} k_{12}(x_{L,1} - x_{C,1})^2 + \frac{1}{2} k_{23} (x_{C,N_C} - x_{R,1})^2.
\end{equation}
Here,
\begin{equation}\label{Halpha_1D}
  H_\alpha = \sum_{i=1}^{N_\alpha} \frac{1}{2} m_\alpha \dot{x}^2_{\alpha,i} + \sum_{i=1}^{N_\alpha-1}\frac{1}{2} k_{\alpha} (x_{\alpha,i} - x_{\alpha,i+1})^2,
\end{equation}
where $x_{\alpha,i}$ is the relative displacement of $i$th atom in part $\alpha$. $N_\alpha$ is the number of atoms in part $\alpha$. Note that for the semi-infinite leads, $N_L$ and $N_R$ are infinite.

The simplest model of the one-dimensional atomic chain is the single-junction case, namely, two semi-infinite leads are directly connected by a spring with a constant strength $k_{12}$. For the Hamiltonian of the single-junction case, we can set $k_{23}=0$, $N_C=0$, and replace $x_{C,1}$ by $x_{R,1}$ in Eq.~(\ref{H_1Dchain}). Within the NEGF approach, the transmission coefficient can be given by \cite{Zhang_2011},
\begin{equation}\label{T_1D}
  \Xi(\omega) = \frac{B^2C_1C_2}{|A_1A_2 - B^2|^2}.
\end{equation}
Here,
\begin{eqnarray}\label{T_1D_par}
  A_i &=& \omega^2 - \frac{k_i}{m_i} (1- \lambda_i) - \frac{k_{12}}{m_i}, \\
  B &=& \frac{k_12}{\sqrt{m_1m_2}},\\
  C_i &=& \frac{\omega}{m_i} \sqrt{4 k_i m_i - \omega^2 m_i^2},
\end{eqnarray}
where $\lambda_i = e^{iq_ia_i}$ with $q_i$ the wave vector and $a_i$ the interatomic spacing.

\begin{figure}
  \includegraphics[width=3.25in]{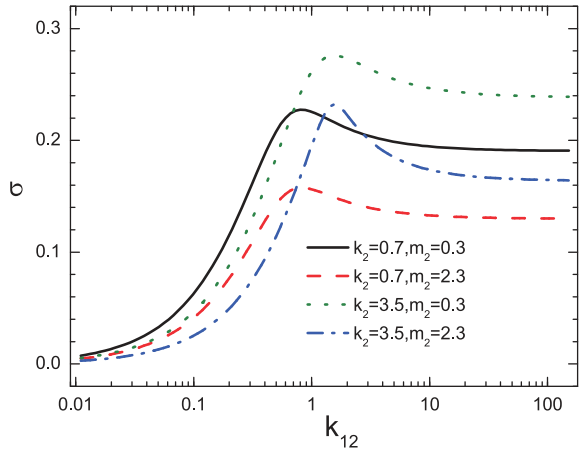}\\
  \caption{Thermal conductance $\sigma$  as a function of interface coupling $k_{12}$ in the single-junction model. Here $k_1 = 1.0$ and $m_1 = 1.0$. Reproduced with permission from Ref.~[\onlinecite{Zhang_2011}].}
  \label{ITC}
\end{figure}

Figure~\ref{ITC} presents the thermal conductance as a function of interfacial coupling $k_{12}$ in the single-junction model. It is found that the thermal conductance initially increases with the increasing interfacial coupling $k_{12}$ and reaches a maximum value. It then decreases slightly and finally approaches a constant value. Zhang et al. found that the maximum thermal conductance occurs when the interface spring equals the harmonic average of the spring constants in two semi-infinite leads, namely, $k_{12}$ satisfies \cite{Zhang_2011}
\begin{equation}\label{k12}
  k_{12} = k_{12,m} = \frac{2k_1k_2}{k_1 + k_2}.
\end{equation}

Besides, the effect of impurity mass and mechanical adhesion on phonon transport was investigated by Saltonstall et al. by introducing an impurity mass and variable bonding into the single-junction model \cite{Saltonstall}. For the case of interface mass, it is found the maximum transmission occurs when the interface mass equals the arithmetic mean of the mass on either side of the interface. For the case of the interface spring, one can maximize the transmission when the interface spring is set to the harmonic mean of the spring constants in two semi-infinite leads, namely, $k_{12,m}$.

The single-junction model can be extended to the two-junction model which involves a central part. In the two-junction model, the transmission wave is scattered by two boundaries, which results in multiple reflections. The transmission behavior can be considered as the combination of the transmission in the single-junction model and the oscillatory behavior due to the multiple scattering. For the two-junction model with homogenous mass and coupling in the central part, it is found that the phonon transmission oscillates with frequency in the envelope lines of minimum and maximum transmission which can be determined by the single-junction model \cite{Zhang_2011}. The interfacial thermal conductance of two-junction model for various mass-graded and coupling-graded materials was investigated by Xiong et al.\cite{Xiong_2020} The optimized homogenous coupler\cite{Chen_2015}, the arithmetic mass-graded and coupling-graded coupler, the geometric mass-graded and coupling-graded coupler, and the coupler with both geometric graded mass and coupling were studied. Relative to the optimized homogenous couplers, the mass-graded or coupling-graded structures were found to be applicable to improve the interfacial thermal conductance of two lead materials with both mismatched impedance and mismatched cutoff frequencies\cite{Xiong_2020}. For the couplers with both geometric graded mass and geometric graded coupling, the interfacial thermal conductance can be maximum enhanced nearly up to sixfold compared to the optimized homogenous case. They also found that the interfacial thermal conductance decreases with the increasing cutoff frequency ratios for all six cases due to the increasing mismatch of the cutoff frequency\cite{Xiong_2020}.

In the above, we discuss the interfacial thermal transport in one-dimensional atomic chains with only linear coupling interactions. However, the nonlinear effect at the interface is another crucial issue for further understanding the fundamental physical mechanism of phonon transport. Zhang et al. introduced a fourth-order nonlinear interaction into the one-dimensional atomic chain model and studied the thermal transport through a solid-solid interface\cite{Zhang_2013}. By using the quantum self-consistent mean-field theory based on the NEGF method, they found that the nonlinear interaction $\lambda$ plays a role to modulate the interfacial linear coupling $k_{12}$ and the effective interfacial coupling can be given by\cite{Zhang_2013}
\begin{equation}\label{k_eff}
  k_{12,eff} = k_{12} + 3 \lambda \left( \frac{\langle u^2_1\rangle}{m_1} - 2\frac{\langle u_1 u_2\rangle}{\sqrt{m_1m_2}} + \frac{\langle u^2_2\rangle}{m_2} \right).
\end{equation}
It was also found that in the weak-interfacial-coupling regime, the interfacial thermal transport is enhanced by the nonlinearity, while the enhancement vanishes in the strong-interfacial-coupling regime.

\begin{figure}
  \includegraphics[width=3in]{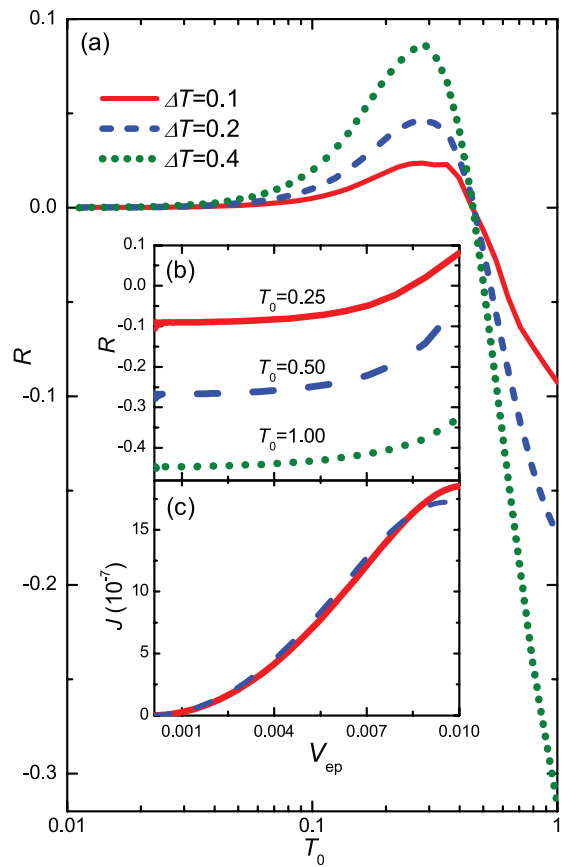}\\
  \caption{(a) Thermal rectification $R$ of the metal-insulator interface as a function of temperature $T_0$ for different temperature gradients with a electron-phonon interaction $V_{ep} = 0.01$. (b) Thermal rectification $R$ as a function of electron-phonon interaction $V_{ep}$ at different temperatures. (c) Thermal current in the forward (solid line) and backward (dashed line) transport a function of electron-phonon interaction $V_{ep}$ at $T_0 = 0.25$. Reproduced with permission from Ref.~[\onlinecite{Zhang_ep}].}
  \label{ep}
\end{figure}

The phonon transport with the weak electron-phonon interaction was also studied in one-dimensional atomic chains. Based on the NEGF method. L\"{u} et al. derived the electrical and energy current of the coupled electron-phonon system by introducing the electron-phonon interaction within the adiabatic Born-Oppenheimer approximation \cite{Lu_ep}. They showed that the self-consistent Born approximation fulfills the electrical and energy current conservation. Zhang et al. studied the thermal conductance and thermal rectification across the metal-insulator interface with electron-phonon interaction by using the NEGF method \cite{Zhang_ep}. They found the thermal conductance has a nonmonotonic behavior as a function of the average temperature of both phonon leads. Moreover, by considering the same temperature of left and right phonon leads and setting $k_{12} = 0$, the phonon contribution in metal was excluded to avoid divergence. Figure~\ref{ep}(a) presents that the thermal rectification changes its sign with the increase of temperature at a relatively larger electron-phonon interaction. While the thermal rectification remains negative at a very weak electron-phonon interaction, as shown in Fig.~\ref{ep}(b). The reverse of thermal rectification can be explained by the relation of thermal currents in the forward and backward directions. At a weak electron-phonon interaction, the forward thermal current is smaller than the backward one, which results in the negative thermal rectification, as presented in Fig.~\ref{ep}(c). When the electron-phonon interaction is strong, the forward thermal current becomes larger than the backward one since more electrons far away from the Fermi surface contribute to the thermal energy, leading to the positive thermal rectification.

Besides, the interfacial thermal transport was studied across anharmonic systems via the one-dimensional atomic chain model. He et al. developed a quantum self-consistent approach to renormalize the anharmonic Hamiltonian to an effective harmonic one, which was used to calculate the interfacial phonon transport within the framework of NEGF method\cite{HeDH}. Fang et al. studied the anharmonic phonon transport across interfaces in nonlinear one-dimensional lattice chains based on the equilibrium MD simulation. An efficient method to calculate the frequency-dependent anharmonic phonon transmission coefficients was proposed based on the linear response theory \cite{Fang2020}.

\begin{figure}
  \includegraphics[width=3.5in]{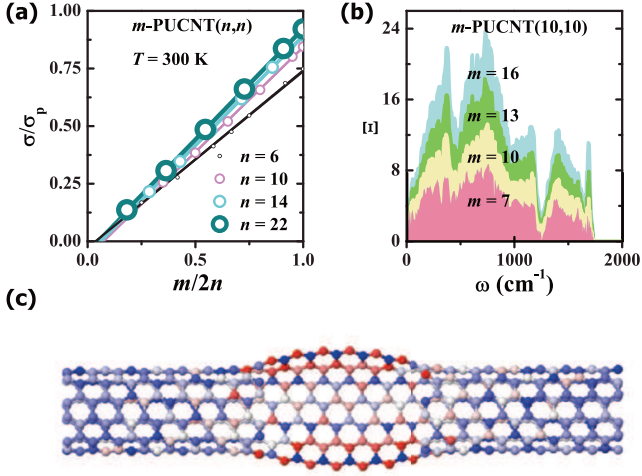}\\
  \caption{(a) Scaled thermal conductance at 300~K of $m$-PUCNT($n$,$n$) as a function of the scaled width $m/2n$. The scaled thermal conductance is defined as the ration of thermal conductance of $m$-PUCNT($n$,$n$) to the thermal conductance of a pristine ($n$,$n$) carbon nanotube. $m$ is the number of zigzag carbon atom chains in the unzipped part. (b) Phonon transmissions of $m$-PUCNT(10,10) as a function of phonon frequency. The highest (lowest) value is represented by red (blue) color. (c) Phonon local density of states of a 7-PUCNT(6,6) at $\omega = 1000$~cm$^{-1}$. Reproduced with permission from Ref.~[\onlinecite{Chen155438}].}
  \label{nanotube}
\end{figure}

Recently, interfacial phonon transports have been extensively studied across the interfaces based on various nanostructures such as single-molecule junctions\cite{Pauly2016,Pauly2017,Cui092201}, self-assembled monolayer interfaces\cite{Hu2010,Lu9b12006,Fan0c02753}, one-dimensional nanotube junctions\cite{Chen155438,Zhang195429}, and two-dimensional heterojunctions\cite{Xu2009,Xu2010,Ding2016,Sadasivam,Zhang445703}. These studies on thermal conductance through actual interfaces confirms the general rules obtained from the NEGF method in the one-dimensional atomic chains. Hu et al. \cite{Hu2010} investigated the phonon transport across a self-assembled monolayer of alkanethiol molecules sandwiched between gold and silicon substrates using the MD simulation. They found that the transmission coefficients exhibit strong and oscillatory dependence on frequency, which agrees with the phonon transmission behavior in the two-junction model\cite{Zhang_2011}. The interfacial thermal conductance of partially unzipped carbon nanotubes was studied by using the NEGF method \cite{Chen155438}. The armchair carbon nanotube was longitudinally unzipped to obtain curved zigzag graphene nanoribbons in its central part, as shown in Fig.~\ref{nanotube}(c). In Fig.~\ref{nanotube}(a), Chen et al. presented that the thermal conductance exhibits a linear dependence on the width of the unzipped graphene nanoribbon region. This can be explained by the enhanced phonon transport channels of carbon nanotubes with a wider width of the unzipped region from the phonon transmission of partially unzipped carbon nanotubes (PUCNTs) shown in Fig.~\ref{nanotube}(b). Such a linear behavior of the thermal conductance to the width of the unzipped graphene nanoribbon region implies that the key factor determined the phonon conduction is the width of the central part.

\subsection{Multi-lead systems}
In Eqs.~(\ref{IMW}) and (\ref{Ifinal}), the thermal currents of systems with two leads are derived. These formulas can be used in the same form for systems with multiple leads when there are no interactions between leads. Similar to the theory of B\"{u}ttiker on the electronic transport in systems with multiple leads, the thermal current flowing out the $\alpha$ lead can be given by \cite{multi,multi1,multi2}
\begin{equation}\label{Imulti}
  J_\alpha = \int^{+\infty}_{0} \frac{d\omega}{2\pi} \omega \sum_{\beta \neq \alpha}  \Xi_{\beta \alpha}(\omega) (n_{\alpha} - n_{\beta}),
\end{equation}
where
\begin{equation}\label{Tmulti}
 \Xi_{\beta \alpha}(\omega) = \mathrm{Tr} (G^r \Lambda_\alpha G^a \Lambda_\beta),
\end{equation}
is the transmission coefficient between the $\alpha$ and $\beta$ leads.

The ballistic thermal transport in three-terminal junctions was studied by Zhang et al. in which the thermal current of the third lead is set to be zero by adjusting its bath temperature \cite{Zhang100301}. The thermal rectification is found in asymmetric three-terminal junctions due to the incoherent phonon scattering from the control lead. By introducing the spin-phonon interaction, the thermal rectification can be found in symmetric three-terminal junctions with an external magnetic field. The ballistic thermal rectification effect was also studied analytically and numerically in asymmetric three-terminal mesoscopic dielectric systems \cite{Ming_2010}. The model of three-terminal junctions is widely extended to study the thermal transport in various two-dimensional nanomaterials\cite{ouyang,Xie2012,GU2020818}. For instance, Ouyang et al. studied the phonon rectification effect of asymmetric three-terminal graphene nanojunctions \cite{ouyang}. They found that the rectification efficiency is strongly dependent on the asymmetry of graphene nanojunctions, which can be significantly improved by increasing the width difference between left and right leads. Moreover, the mode-dependent phonon transport in three-terminal graphene nanojunctions was investigated by Gu et al. based on the NEGF method and the acoustic modes were found to contribute higher transmission coefficients between the zigzag graphene nanoribbon and the third lead\cite{GU2020818}.

\begin{figure}
  \includegraphics[width=3.25in]{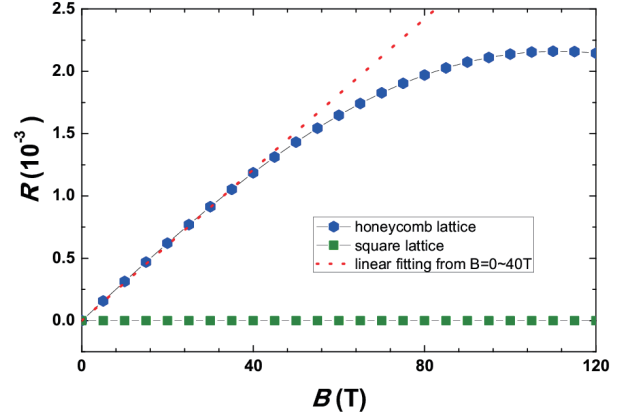}\\
  \caption{Hall temperature difference $R$ as a function of magnetic filed $B$ at temperature $T=5.45$~K. Reproduced with permission from Ref.~[\onlinecite{Zhang_2009}].}
  \label{PHE}
\end{figure}

In electronic transport, four-terminal devices have been extensively used to study the spin Hall effect for two-dimensional mesoscopic systems in which a transverse charge accumulation is induced by a longitudinal electric field \cite{Xing2006,Xing2007,Wei2020}. Analogous to the electric Hall effect, the phonon Hall effect where a transverse heat flow in dielectrics is induced by a longitudinal temperature difference has been discovered experimentally in 2005 \cite{Strohm}. Using the NEGF approach, Zhang et al. studied the phonon Hall effect for paramagnetic dielectrics in four-terminal nanojunctions.\cite{Zhang_2009} Fig.~\ref{PHE} presents the Hall temperature difference for the honeycomb and square lattices with nearest-neighbor couplings under different magnetic filed at the temperature of $T=5.45$~K. For the honeycomb lattice, it is found that the Hall temperature difference exhibits the linear relation to the magnetic field lesser than 40~T. The fitted slope is about $3\times 10^{-5}$~K~T$^{-1}$, which is comparable to the experimental results\cite{Strohm}. When the magnetic field is extremely large, the Hall temperature difference decreases slightly with the increasing magnetic field. However, the phonon Hall effect can not be obtained in the square lattice with nearest-neighbor couplings due to the mirror reflection symmetry of the dynamic matrix. Once the next-neighbor couplings is considered in the square lattice, the phonon Hall effect can then be obtained.

\subsection{Time-dependent phonon transport in the transient regime}
In the past decade, most of the theoretical works on thermal transport focus on the calculation of steady-state phenomena. However, the time-dependent phonon current in the transient regime is also an important question. Recently, the transient phonon transport was studied in arbitrary harmonic systems connected to phonon baths by abruptly turning on the coupling between leads within the partition scheme based on the NEGF method \cite{JsWang2010,Tuovinen}.

Considering a single-junction one-dimensional chain model in which the left and right leads are initially uncoupled. Before $t=0$, it is assumed that the left and right leads are in thermal equilibrium with temperature $T_L$ and $T_R$, respectively. The coupling between the left and right leads is suddenly switched on at $t=0$ by an interparticle harmonic potential with a spring constant $k$. The time-dependent phonon current in the transient regime can be expressed as \cite{JsWang2010}
\begin{equation}\label{Jt}
 J_L(t) = k \mathrm{Im} \left[ \frac{\partial G^{RL,<}(t_1,t_2)}{\partial t_2} \right]_{t_1=t_2=t}.
\end{equation}
Here, the time-derivative of $G^{RL,<}(t_1,t_2)$ is given by \cite{JsWang2010}
\begin{eqnarray}\label{Grlt}
 &&\frac{\partial G^{RL,<}(t_1,t_2)}{\partial t_2}  \nonumber\\
 &=& - k \int^t_0 dt_a G^{RL,r}(t_1,t_a) \frac{\partial G^{RL,<}_1(t_a,t_2)}{\partial t_2} \nonumber\\
 && - k \int^t_0 dt_a G^{RL,<}_1(t_1,t_a) \frac{\partial G^{RL,a}(t_a,t_2)}{\partial t_2} \nonumber\\
 && + k^2 \int^t_0 dt_a \int^t_0 dt_b G^{RL,r}(t_1,t_a) G^{RL,<}_1(t_a,t_b) \nonumber\\
 && \times \frac{\partial G^{RL,a}(t_b,t_2)}{\partial t_2} +\frac{\partial G^{RL,<}_1(t_1,t_2)}{\partial t_2},
\end{eqnarray}
where
\begin{eqnarray}\label{Grlt1}
G^{RL,<}_1(t_1,t_2) = -k \int^t_0 dt_a \Big[ g^{R,r}(t_1-t_a) g^{L,<}(t_a-t_2) \nonumber\\
+ g^{R,<}(t_1-t_a) g^{L,a}(t_a-t_2) \Big],  \nonumber\\
\end{eqnarray}
and
\begin{eqnarray}\label{Grlt2}
G^{RL,\beta}(t_1,t_2) &=& -k \int^t_0 dt_a G^{RL,\beta}_1(t_1,t_a) G^{RL,\beta}(t_a,t_2) \nonumber\\
&& + G^{RL,\beta}_1(t_1,t_2),
\end{eqnarray}
with $\beta = r, a$. The first-order term of Eq.~(\ref{Grlt2}) can be expressed as
\begin{equation}\label{Grlt3}
G^{RL,\beta}_1(t_1,t_2) = -k \int^t_0 dt_a g^{R,\beta}(t_1-t_a) g^{L,\beta}(t_a-t_2).
\end{equation}

In order to calculate the time-dependent phonon current, the time variable is discretized into a large numbers of segments. Since the analytic expressions for the equilibrium surface Green's function $g^{L,\gamma}$ and $g^{R,\gamma}$ ($\gamma = r, a, <$) in Eqs.~(\ref{Grlt1}) and (\ref{Grlt3}) have been given in the frequency domain \cite{NEGF4}, the corresponding time-dependent surface Green's functions can be numerically calculated by Fourier transform to obtain $G^{RL,\gamma}_1(t_1,t_2)$. Then one can solve $G^{RL,r}(t_1,t_a)$ and $\frac{\partial G^{RL,a}(t_a,t_2)}{\partial t_2}$ required in Eq.~(\ref{Grlt}) from Eq.~(\ref{Grlt2}) by transforming the integral into a sum. Finally, by solving the time-derivative of the Green's function $G^{RL,<}(t_1,t_2)$ in Eq.~(\ref{Grlt}), the time-dependent phonon current in the transient regime can be calculated.

\begin{figure}
  \includegraphics[width=3.25in]{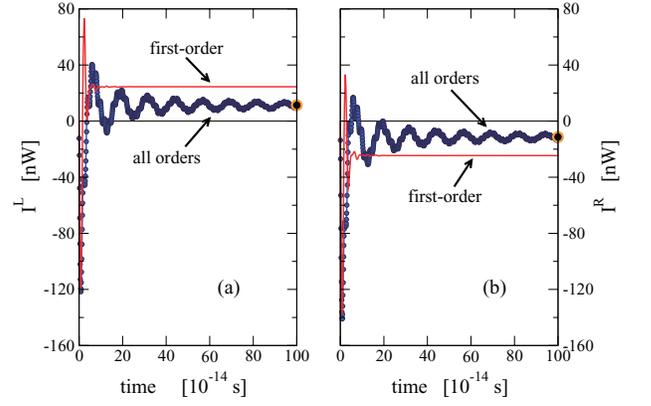}\\
  \caption{Time-dependent phonon current flowing out of the (a) left and (b) right leads in the transient regime. The red lines are the results when only the first-order term in the perturbation is used in the calculation. The temperatures of the left and right leads are set to be $T_L=330$~K and $T_R=270$~K, respectively. Reproduced with permission from Ref.~[\onlinecite{Wang019902}].}
  \label{transientJ}
\end{figure}

Figure~\ref{transientJ} plots the time-dependent phonon current in the transient regime by setting the temperatures of left and right leads to be $T_L=330$~K and $T_R=270$~K, respectively. Once the coupling between left and right leads is switched on, the transient currents of both leads flow in an unexpected direction, namely, flow from the colder lead to the hotter one. The transient currents then increase to positive and gradually approach the steady-state that can be calculated directly from the Landauer formula in the long-time limit. The time-dependent currents exhibit oscillatory behavior and the oscillation frequency is comparable to the highest phonon frequencies available in the system.

In addition, the transient behavior of time-dependent phonon current can also be studied by the full-counting statistics of heat transport in harmonic junctions based on the NEGF technique \cite{Wangjs2011,Agarwalla,Agarwalla2015}. Wang et al. derived the generating function of energy counting statistics for phononic junctions which can be expressed in terms of contour-ordered Green's function as \cite{Wangjs2011},
\begin{equation}\label{CGF}
\ln Z(\xi) =  - \frac{1}{2} \mathrm{Tr}_{j,\tau} \ln (1 - G \Pi^A).
\end{equation}
Here, the notation $\mathrm{Tr}_{j,\tau}$ represents the trace in both space index $j$ and contour time $\tau$. $G$ is the Green's function defined on the Keldysh contour and $\Pi^A$ is obtained from the difference of the original lead self-energy and the lead energy shifted by the contour time arguments. In the long-time limit, the cumulant generating function for large $t_M$ can be expressed using Green's functions in the frequency domain,
\begin{eqnarray}\label{CGF2}
\ln Z(\xi) &=&  -t_M \int^{+\infty}_{-\infty} \frac{d\omega}{4\pi} \ln \det \{  1 - G^r \Pi_L G^a \Pi_R [(e^{i\xi \omega}-1) n_L \nonumber\\
&& + (e^{-i\xi \omega}-1) n_R + (e^{i\xi \omega} - e^{-i\xi \omega} -2) n_L n_R] \}.
\end{eqnarray}
This formalism is first given by Saito and Dhar \cite{Saito180601,Saito041121} and satisfies the steady-state fluctuation theorem. Agarwalla et al. then investigated the full counting statistics of heat transferred in harmonic chains in the presence of both temperature gradients and time-dependent driving forces\cite{Agarwalla}. The cumulant generating function for heat transferred from the leads to the central region was calculated based on the two-time measurement concept using the NEGF method. The transient behavior and steady-state fluctuations were studied in atomic chains with different initial conditions and the results were generalized for systems with multiple heat baths.

\section{Thermoelectric transport}\label{sec3}
\subsection{dc thermoelectric transport}
Since the observation of the Seebeck effect which revealed the interplay between thermal gradient and electric potential, thermoelectricity has attracted much attention due to its potential applications in power generation and refrigeration. Recently, the Seebeck effect was studied in various nanostructures which provides new opportunities for designing thermoelectric devices with high $ZT$ values \cite{Dubi_RMP,Hochbaum,Reddy1568,Gunst2011,Chen2010,Yang2012,Xing235411,Wei245432,Wang2014,Wang2016,Zhou2017}. The Seebeck coefficient has been successfully measured in molecular junctions by trapping molecules between two gold electrodes, which offers a promising way to study the fundamental physics in thermoelectric energy conservation\cite{Reddy1568}. A significant $ZT$ value of 0.6 is achieved experimentally at room temperature in one-dimensional silicon nanowires with rough surfaces which exhibits a 100-fold reduction of thermal conductivity due to the efficient phonon scattering\cite{Hochbaum}.

In dc transport, the thermopower is related to the electric conductance of nanodevices which can be simply modeled by the well-known Landauer-B\"{u}ttiker formalism within the NEGF approach. Similar to the phonon energy current, the electric current and the electric heat current for spin-degenerate systems can be given by ($\hbar=e=1$ for simplicity) \cite{Butcher,Sivan}
\begin{equation}\label{Ie}
  I = \int^{+\infty}_{-\infty} \frac{d\epsilon}{\pi} \mathcal{T}(\epsilon) (f_{L} - f_{R}),
\end{equation}
and
\begin{equation}\label{Ie}
  I^h = \int^{+\infty}_{-\infty} \frac{d\epsilon}{\pi} (\epsilon-\mu) \mathcal{T}(\epsilon) (f_{L} - f_{R}).
\end{equation}
Here, $f_{\alpha}(\epsilon) = 1/[\exp(\frac{\epsilon-\mu_\alpha}{T_\alpha})+1]$ is the Fermi-Dirac distribution function with $\mu_\alpha$ the chemical potential in lead $\alpha$ ($k_B$ = 1 for simplicity). $\mathcal{T}(E)$ is the electric transmission spectrum
\begin{equation}\label{Tele}
  \mathcal{T}(E) = \mathrm{Tr} (\mathcal{G}^r \Gamma_L \mathcal{G}^a \Gamma_R),
\end{equation}
where $\Gamma_\alpha$ and $\mathcal{G}^{r(a)}$ are the electric bandwidth function of lead $\alpha$ and the electric retarded (advanced) Green's function, respectively.

In the linear response, namely, under small bias voltage and small temperature gradient, the electric current can be linearly expanded \cite{Mahanbook}
\begin{eqnarray}\label{Iexpand}
  I &=& \int^{+\infty}_{-\infty} \frac{d\epsilon}{\pi} \mathcal{T}(\epsilon)\left[ - \frac{\partial f}{\partial \epsilon} \Delta V - \frac{\partial f}{\partial \epsilon} \left( \frac{\epsilon - \mu}{T}\right) \Delta T \right] \nonumber\\
  & \equiv & L_0 \Delta V + \frac{L_1}{T} \Delta T.
\end{eqnarray}
Here, $\Delta V = V_L - V_R$ and $\Delta T = T_L - T_R$ are the bias difference and temperature difference between the left and right leads, respectively, and
\begin{equation}\label{Ln}
 L^n = - \int^{+\infty}_{-\infty} \frac{d\epsilon}{\pi} \mathcal{T}(\epsilon) (\epsilon - \mu)^n \left( \frac{\partial f}{\partial \epsilon} \right).
\end{equation}
Similarly, the electric heat current can be expanded as,
\begin{equation}\label{Ihexpand}
 I^h = L_1 \Delta V + \frac{L_2}{T} \Delta T.
\end{equation}

The Seebeck coefficient, also called thermopower, which measures the magnitude of $\Delta V$ to balance the electric current along the reverse direction due to $\Delta T$, is defined as,
\begin{equation}\label{Sdef}
 S = - \frac{\Delta V}{\Delta T} \Bigg|_{I=0}.
\end{equation}
From Eq.~(\ref{Iexpand}), it is easy to obtain,
\begin{equation}\label{Seebeck}
 S = - \frac{1}{T} \frac{L_1}{L_0}.
\end{equation}

We can also define the thermal conductance of electrons $\kappa_{el}$ when the electric current is zero. From $I=0$ and $I^h = \kappa_{el} \Delta T$, we obtain
\begin{equation}\label{kel}
 \kappa_{el} = \frac{1}{T} \left( {L_2 - \frac{L_1^2}{L_0} } \right).
\end{equation}
By further defining the electron conductance from Ohm's law $\mathbb{G} = I/ \Delta V = L_0$, we can replace $L_n$ in Eqs.~(\ref{Iexpand}) and (\ref{Ihexpand}) by $\mathbb{G}$, $S$, and $\kappa_{el}$ and obtain the following relation \cite{ChenXB}
\begin{equation}\label{IIhrelation}
\left(
  \begin{array}{c}
    I \\
    I^h \\
  \end{array}
\right) =
\left(
  \begin{array}{cc}
    \mathbb{G} & \mathbb{G}S \\
    \mathbb{G}ST & \kappa_{el} +  \mathbb{G}S^2 T \\
  \end{array}
\right)
\left(
  \begin{array}{c}
    \Delta V \\
    \Delta T \\
  \end{array}
\right).
\end{equation}

Moreover, we can describe the thermoelectric effect by the figure of merit, $ZT$, which gives the maximum efficiency of energy conservation in thermoelectric devices. It can be calculated by
\begin{equation}\label{ZT}
 ZT = \frac{\mathbb{G}S^2}{\kappa_{el}+\kappa_{ph}} T.
\end{equation}

Based on the NEGF method, the thermoelectric properties were extensively investigated in low-dimensional nanostructures \cite{Chen2010,Gunst2011,Yang2012,Xing235411,Wei245432}. Gunst et al. studied the thermoelectric properties of graphene antidot structures by using the $\pi$-tight-binding model. They found that the $ZT$ can exceed 0.25 at room temperature and it is highly sensitive to the structure of antidot edges\cite{Gunst2011}. Chen et al. studied the thermoelectric properties of graphene nanoribbons, junctions, and superlattices \cite{Chen2010}. Their findings indicate that the thermoelectric behavior is controlled by the width of the narrower part of graphene junctions. Moreover, the thermoelectric transport was studied in hybrid graphene and boron nitride nanoribbons and it was found that the $ZT$ value can be significantly enhanced by periodically embedding hexagonal boron nitride into graphene nanoribbons\cite{Yang2012}. Besides, the effect of electron-phonon coupling and electron-electron interaction on thermoelectric transport was studied in a single molecular junction and it was found that $ZT$ can be enhanced by increasing electron-phonon coupling and Coulomb repulsion\cite{Ren2012}.

In 2008, the spin Seebeck effect which generates the spin voltage from temperature gradient has been observed experimentally in a metallic magnet by Uchida et al \cite{RN109}. How to manipulating and control the spin degrees of freedom in thermal ways has attracted much attention. Spin caloritronics concerning coupled spin, charge, and energy transport in magnetic structures was introduced to focus on the relations between spin and heat current\cite{Bauer2010,Bauer2012}. In spin caloritronics, various nonequilibrium phenomenons driven by thermal gradient have been investigated such as thermal spin transfer torque\cite{Hatami,Zhang064414}, spin-polarized currents\cite{Zeng2000049,Wang2016}, and pure spin currents \cite{Yu246601,Cheng045302,Wang2018}. Using the first-principles calculation combined with the NEGF method, a strongly spin-polarized current due to temperature difference was obtained in magnetized zigzag graphene nanoribbons by breaking the electron-hole symmetry\cite{Zeng2000049}. The spin current can be completely polarized by tuning the gate voltage. Moreover, a pure spin current was generated in a triangulene-based molecular junction on a large scale by changing the temperature gradient and gate voltage \cite{Wang2018}.

Apart from the charge and spin degrees of freedom, the valley degree of freedom can be used in valleytronics for the application of information processing similar to spin used in spintronics \cite{Xiao,Nebel,Rycerz,White,Chang}. A complete valley polarized electronic current has been obtained by simply introducing the line defect in graphene \cite{White}. The generation of a pure bulk valley current without net charge current through quantum pumping has also been reported in graphene by using the well-known Dirac Hamiltonian \cite{Chang}. Analogous to spin caloritronics, valley caloritronics, a combination of valleytronics and thermoelectrics, has been proposed to generate a valley polarized current or a pure valley current using thermal means \cite{Yu2016451,ZHANG2018183,Chen2015}.

\begin{figure}
  \includegraphics[width=3.25in]{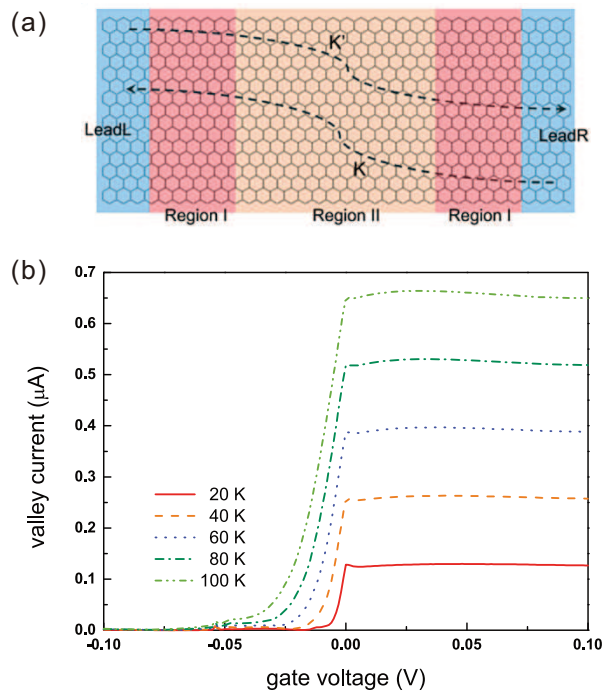}\\
  \caption{(a) Schematic diagram of zigzag graphene nanoribbons with two semi-infinity leads (blue shadow). Two static gate regions with $v_{g1} = 0.5$~V (red shadow) and $v_{g2}$ (orange shadow) is tunable in the central region. (b) Valley current as a function of $v_{g2}$ under different temperature gradients with fixed $T_R = 0$~K. Reproduced with permission from Ref.~[\onlinecite{Yu2016451}].}
  \label{valley}
\end{figure}

The valley Seebeck effect was first proposed in gate tunable zigzag graphene nanoribbons by Yu et al. using the tight-binding model within the NEGF framework \cite{Yu2016451}, as shown in Fig.~\ref{valley}(a). From the unique band structure of zigzag graphene nanoribbons, one can find that the momentum and valley index of electrons in the first subband are locked together. Therefore, the left- and right-moving electrons have valley index $K$ and $K'$, respectively. Since at given energy the sign of $f_L - f_R$ determines the direction of electron flow and the valley index, the valley current of zigzag graphene nanoribbons can be simply expressed as \cite{Yu2016451},
\begin{equation}\label{Ivalley}
 I_v = \int \frac{dE}{2\pi} \mathrm{sgn}(f_L-f_R) (f_L-f_R) \mathcal{T}(E).
\end{equation}

A pure valley current can be generated by the thermal gradient as well as the external bias. In order to control the pure valley current, the gate voltage $v_{g2}$ applied in the central region is modulated. Fig.~\ref{valley}(b) presents the pure valley current as a function of $v_{g2}$ at different temperature gradient with $T_R = 0$~K. It is found there is a threshold gate voltage to open the valley current. Both the threshold gate voltage and on valley current are proportional to the temperature gradient and the valley current reaches the maximum value at the neutral gate voltage. These behaviors suggest the potential applications as a valley field-effect transistor driven by the temperature gradient.

Moreover, the dephasing effect and doping effect on the valley Seebeck effect in zigzag graphene nanoribbons were studied\cite{Zhang064414}. It was found that the dephasing effect only reduces the magnitude of pure valley current. While the valley polarized current occurs by random doping of boron and nitrogen atoms and the valley polarization can be effectively tuned by the doping concentration. Both the valley polarized current and pure valley current can also be obtained in wedge-shaped zigzag graphene nanoribbon junctions\cite{Chen2015}. In addition to graphene-based nanostructures, valley and spin thermoelectric transport has also been investigated in silicene junctions \cite{Zhai245405,Zhi2014} and group-IV monolayers\cite{Zhai2017}.

\subsection{time-dependent thermoelectric transport in the transient regime}
Besides the static thermoelectric behavior, time-dependent thermoelectric transport is also an important issue that may provide fundamental insights to understand the thermal response of mesoscopic systems. Generally, there are two different schemes to study time-dependent quantum electronic transport. One is the partition-free scheme (Cini scheme) in which the initial state of the system is assumed to be at equilibrium that can be described by a thermal density matrix\cite{Stefanucci,Cini}. Then the system can be perturbed by applying a time-dependent voltage bias. Another way is the partitioned scheme (Caroli scheme) which assumes that the two-probe system is disconnected initially and the coupling between the scattering region and two leads is treated as the time-dependent perturbation \cite{Caroli_1971,Caroli2}. In the following, we will discuss the time-dependent thermoelectric transport in the transient regime using the NEGF method within the Caroli scheme.

Within the Caroli scheme, the leads are assumed to at equilibrium states with the temperature $T_\alpha$ and applied bias $V_\alpha$ before $t=0$ and the couplings between leads and the central region are turned on at $t=0$. The exact solution of the transient electric current that beyond wide-band limit (WBL) can be given by \cite{Yu_2020}
\begin{equation}\label{currentA}
I_L(t)=\int \frac{d\epsilon}{2\pi} {\rm Tr} [A(\epsilon ,t){\Sigma}^{<} (\epsilon)B_{L}(\epsilon ,t)+A(\epsilon ,t){\Sigma}^{<}_L (\epsilon)] +h.c.
\end{equation}
where $A$ is the spectral function defined as
\begin{equation} \label{spectral}
A(\epsilon,t)=\int^{t}_0 dt' \mathcal{G}^r(t,t')e^{i\epsilon(t-t')},
\end{equation}
which can be expressed by Green's functions in energy domain within the Caroli scheme,
\begin{equation} \label{Afinal}
A(\epsilon,t) = \mathcal{G}^r(\epsilon) + \int \frac{d\epsilon'}{2\pi i} \frac{e^{-i(\epsilon'-\epsilon)t}}{\epsilon -\epsilon' + i0^+} \mathcal{G}^r(\epsilon'),
\end{equation}
and
\begin{equation}\label{B}
B_{L}(\epsilon ,t)=\int \frac{d\epsilon'}{-2\pi i} \mathcal{G}^a(\epsilon') {\Sigma}^{a}_L (\epsilon') \frac{e^{i(\epsilon' -\epsilon)t}}{\epsilon -\epsilon' +i0^+}.
\end{equation}

In order to study the transient thermoelectric transport, the applied biases of left and right leads are assumed to be $V_L = \Delta V$ and $V_R = 0$, respectively. A temperature difference $\Delta T$ of two leads are introduced by setting $T_L = T_0 + \Delta T$ and $T_R = T_0$. We can found that $A(\epsilon ,t)$ and $B^{\chi}_{\alpha}(\epsilon ,t)$ only depend on the applied bias while ${\Sigma}^{<}_L (\epsilon)$ depends on both the applied bias $\Delta V$ and the temperature gradient $\Delta T$.

In the linear response regime, the retarded Green's function of electrons in the steady state $\mathcal{G}^r(\epsilon)$ can be expanded to the first order in $\Delta V$ according to the Dyson equation\cite{WangBG},
\begin{equation} \label{Grexpand}
\mathcal{G}^r(\epsilon)= \tilde{\mathcal{G}}^r(\epsilon) - \tilde{\mathcal{G}}^r(\epsilon) \dfrac{\partial \tilde{\Sigma}^r_{L} (\epsilon)}{\partial \epsilon}\tilde{\mathcal{G}}^r(\epsilon) \Delta V.
\end{equation}
Here, the superscript $'\sim'$ is used to denote the quantities in the absence of applied bias and temperature gradient. Similarly, the Fermi-Dirac distribution can be expanded as
\begin{equation} \label{Fexpand}
f_{L} (\epsilon + \Delta V)= \tilde{f} (\epsilon) + \dfrac{\partial \tilde{f}}{\partial \epsilon} \Delta V - \frac{\epsilon}{T} \dfrac{\partial \tilde{f}}{\partial \epsilon} \Delta T ,
\end{equation}
Then, the transient electric current in the left lead can be expressed as,
\begin{equation} \label{IIexpand}
I_L(t)= \tilde{I}_L(t) + \mathcal{G}_{V} (t) \Delta V + \mathcal{G}_{T} (t) \Delta T .
\end{equation}
Here, $\tilde{I}_L(t)$ is the equilibrium transient electric current of the left lead in the absence of voltage gradient and temperature gradient. It is solely contributed from the switching of the coupling between the quantum dot and leads. $\mathcal{G}_{V} (t) $ is the electric conductance of the left lead \cite{Chenj_2015,Yu_2020}
\begin{eqnarray} \label{GV}
\mathcal{G}_{V} (t) &=& \int \frac{d\epsilon}{2\pi} {\rm Tr} [A_V(\epsilon ,t)\tilde{{\Sigma}}^{<} (\epsilon){\tilde{B}}_{L}(\epsilon ,t) \nonumber \\
&&+\tilde{A}(\epsilon ,t){\Sigma}^{<}_{V} (\epsilon){\tilde{B}}_{L}(\epsilon ,t) + \tilde{A}(\epsilon ,t)\tilde{{\Sigma}}^{<} (\epsilon)B_{V}(\epsilon ,t) \nonumber \\
&&+ A_V(\epsilon ,t)\tilde{{\Sigma}}^{<}_L (\epsilon) + \tilde{A}(\epsilon ,t){\Sigma}^{<}_{V} (\epsilon)] +h.c.,
\end{eqnarray}
and $\mathcal{G}_{T}$ is the thermal coefficient due to electrons of the left lead
\begin{eqnarray} \label{GT}
\mathcal{G}_{T} (t) &=& \int \frac{d\epsilon}{2\pi} {\rm Tr} [\tilde{A}(\epsilon ,t){\Sigma}^{<}_{T} (\epsilon)\tilde{B}_{L}(\epsilon ,t) \nonumber \\
&&+\tilde{A}(\epsilon ,t){\Sigma}^{<}_{T} (\epsilon)] + h.c.
\end{eqnarray}
Here,
\begin{equation} \label{AV}
A_V(\epsilon ,t) = \int \frac{d\epsilon'}{2\pi i} \bigg[\tilde{\mathcal{G}}^r(\epsilon') \frac{\partial \tilde{\Sigma}^{r}_{L} (\epsilon')}{\partial \epsilon'}\tilde{\mathcal{G}}^r(\epsilon')\bigg] \dfrac{e^{-i(\epsilon' -\epsilon)t}}{\epsilon-\epsilon'-i0^+} ,
\end{equation}
\begin{eqnarray} \label{BV}
B_{V}(\epsilon ,t) &=& \int \frac{d\epsilon'}{-2\pi i} \bigg[ \tilde{\mathcal{G}}^a(\epsilon') \frac{\partial \tilde{\Sigma}^{a}_{L} (\epsilon')}{\partial \epsilon'}\tilde{\mathcal{G}}^a(\epsilon') \tilde{\Sigma}^{a}_{L} (\epsilon')  \nonumber \\
&& + \tilde{\mathcal{G}}^a(\epsilon') \frac{\partial \tilde{\Sigma}^{a,0}_{L} (\epsilon')}{\partial \epsilon'} \bigg] \frac{e^{i(\epsilon' -\epsilon)t}}{\epsilon-\epsilon'+i0^+} , \nonumber\\
\end{eqnarray}
\begin{equation} \label{sigmaV}
{\Sigma}^{<}_V (\epsilon) = \frac{\partial \tilde{\Sigma}^{<}_{L} (\epsilon)}{\partial \epsilon} ,
\end{equation}
\begin{equation} \label{sigmaT}
{\Sigma}^{<}_T (\epsilon) = i \tilde{\Gamma}_{L} (\epsilon) \frac{\epsilon}{T_0} \frac{\partial \tilde{f}(\epsilon)}{\partial \epsilon},
\end{equation}
where $\tilde{\Gamma}_{L} (\epsilon)$ is the linewidth function of left lead with $\Delta V = 0$.

By setting $\Delta I_L(t) = I_L(t) - \tilde{I}_L(t) = 0$, the time-dependent Seebeck coefficient in the transient regime can be obtained from Eq.~(\ref{IIexpand}),
\begin{equation} \label{Seebeck}
S_{L} (t) = -\frac{\Delta V}{\Delta T} = \frac{\mathcal{G}_{T} (t)}{\mathcal{G}_{V} (t)} .
\end{equation}

\begin{figure}
  \includegraphics[width=3in]{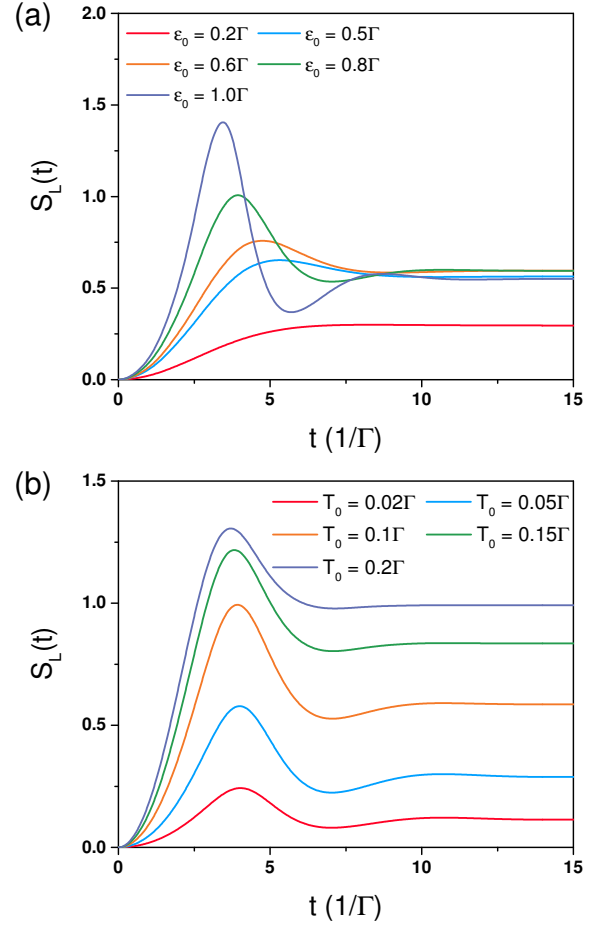}\\
  \caption{(a) Time-dependent Seebeck coefficient in the transient regime with different energy levels of quantum dot $\epsilon_0$. The reference temperature of leads is set to be $T_0 = 0.1\Gamma$. (b) Time-dependent Seebeck coefficient in the transient regime with different reference temperatures $T_0$. The energy level of quantum dot is set to be $\epsilon_0 = 0.5\Gamma$. The bandwidth is set to be $W = 10\Gamma$. Reproduced with permission from Ref.~[\onlinecite{Yu_2020}]. }
  \label{transientS}
\end{figure}

The time-dependent Seebeck coefficient was then studied in the transient regime for a single-level quantum dot with Lorentzian linewidth by Yu et al.\cite{Yu_2020} Fig.~\ref{transientS}(a) presents the transient Seebeck coefficient with different energy levels of the quantum dot under a fixed reference temperature. It was found that the transient Seebeck coefficient oscillates in time and the oscillation frequency is inversely promotional to the energy level of the quantum dot. The time-dependent Seebeck coefficient exhibits a significant enhancement in the transient regime and the enhancement can be improved by the increasing energy level of the quantum dot. The transient Seebeck coefficient can also be enhanced by the reference temperature in the linear response regime and the enhancement grows significantly with the increasing reference temperature, as shown in Fig.~\ref{transientS}(b). These results show the intrinsic damped oscillatory behavior in the time-dependent Seebeck coefficient in the transient regime and the enhancement of transient Seebeck coefficient can be tuned by either the energy level of quantum dot or the reference temperature of leads.

The transient spin current under a thermal switch is also investigated within the partitioned scheme using the NEGF method and an ultrafast enhancement of the spin current in the transient regime is observed\cite{Chen_2018}. Besides the Caroli scheme, the time-dependent charge and heat currents driven by temperature gradients were studied by using the Luttinger-field approach within the Cini scheme in which the Luttinger thermomechanical potential was employed to simulate a sudden change of temperature in leads \cite{Eich2014,Eich2016,Lozej}. An enhanced thermopower was obtained in nanoscale devices under a time-dependent gate voltage by using the NEGF method within the wide-band limit based on the Cini scheme\cite{Adeline}. The time-dependent thermoelectric transport was also studied in multi-terminal noninteracting systems by tight-binding models within a gauge-invariant theoretical framework which is similar to the Caroli scheme \cite{Adel2020}. Moreover, the Caroli and Cini schemes to study the time-dependent transport in mesoscopic systems have been compared by using the NEGF approach and it was proved that the formulas of lesser Green's function and time-dependent electric current obtained by the Cini and Caroli scheme are equivalent \cite{Odashima,Ridley2018}.

In addition to the Seebeck coefficient, the time-dependent thermal transport also focuses on the energy and heat currents driven by external biases as well as temperature gradients in the transient regime. The expressions of transient energy and heat currents in mesoscopic systems were obtained by using the NEGF method with the wide-band limit\cite{Adeline,Lombardo2016}. An exact solution of transient heat current was also derived that goes beyond the wide-band limit and a time-dependent framework to study the transient heat current in realistic nanoscale devices from first principles was proposed \cite{Yu2014}.

Based on the path-integral NEGF method, the full-counting statistic calculations were applied as another way to study the time-dependent energy currents and thermodynamic transport in the transient regime \cite{Yu2016,Li165425,Ridley2019,Tang155430,Tang2018}. The cumulate generating function for full-counting statistics of transferred energy in the transient regime was derived with a two-time measurement scheme by using the Keldysh NEGF approach and the transient behavior and fluctuations of transferred energy were studied for both single- and double-quantum-dot systems\cite{Yu2014}. The thermal rectification and negative differential effects of full-counting statistics, as well as the heat engine performance were investigated in a spin Seebeck engine\cite{Tang2018}. The cumulant generating functions of heat and spin currents were obtained which were demonstrated to obey special fluctuation symmetry relations.

\section{Conclusion}\label{sec4}
In this review, we focus on the thermal transport in mesoscopic systems studied by using the NEGF approach. We first give a brief introduction to the phonon NEGF method and the detailed formalism of phonon current is presented in terms of phonon Green's function. Various theoretical investigations on quantum thermal transport in mesoscopic systems are discussed, which covers the interfacial thermal transport in one-dimensional atomic chains, the effect of nonlinearity and electron-phonon coupling on the interfacial thermal conductance, phonon transport in multi-terminal systems, and time-dependent phonon transport in the transient regime. We also introduce the application of the NEGF method on the thermoelectric transport within the linear response theory. The formalism of the Seebeck coefficient and $ZT$ value in the dc thermoelectric transport are given and they are extended to the spin and valley caloritronics. The time-dependent thermoelectric transport in the transient regime is further discussed within the Caroli scheme.

There are still many issues that deserve future investigation in the field of thermal transport. For instance, manipulating phonons in two- and three-dimensional interfaces to achieve low interfacial thermal conductance, controlling the chirality of phonon in topological insulators, controlling other (quasi) particles such as magnons and skyrmions by the means of thermal, and discovering new materials with low thermal conductance and high electric conductance for optimized thermoelectric performance. From the aspect of the development of the NEGF method, time-dependent thermal transport, higher-order fluctuations of thermal current, and the NEGF-DFT framework for phonon transport, are still open to address. We hope this brief review can inspire more investigations on quantum thermal transport and provide helpful guidance on thermal engineering and applications.

\begin{acknowledgments}
This work was financially supported by the National Natural Science Foundation of China (Grants Nos. 12074190, 11975125, 11890703, and 11874221).
\end{acknowledgments}

\end{document}